**39**

# Event-centric Query Suggestion for Online News

SACHIN, Indian Institute of Technology Roorkee
DHAVAL PATEL, IBM Research, NY

Query suggestion refers to the task of suggesting relevant and related queries to a search engine user to help in query formulation process and to expedite information retrieval with minimum amount of effort. It is highly useful in situations where the search requirements are not well understood and hence it has been widely adopted by search engines to guide users' search activity. For news websites, user queries have a time sensitive nature inherent in them. When some new event happens, there is a sudden burst in queries related to that event and such queries are sustained over a period of time before fading away with that event. In addition to this temporal aspect of search queries fired at news websites, they have an addition distinct quality, i.e., they are intended to get event related information majority of the times. Existing work on generating query suggestions involves analyzing query logs to suggest queries which are relevant and related to the search intent of the user. But in case of news websites, when there is a sudden burst in information related to a particular event, there are not enough search queries fired by other users which leads to lack of click data, and hence giving query suggestions related to some old event or even some irrelevant suggestions altogether. Another problem with query logs in the context of online news is that, they mostly contain queries related to popular events and hence fail to capture less popular events or events which got overshadowed by some other more sensational event.

We propose a novel approach to generate event-centric query suggestions using metadata of news articles published by news media. Our system is based on grouping and ranking of News Keywords in Day wise and Duration wise event clusters. Query suggestions generated by our system are not only representative of all the news events reported by news media, but they also captures the time sensitive aspect of queries fired at a news website. We compared our proposed framework with existing state of the art query suggestion mechanisms provided by Google News, Bing News, Google Search and Bing Search on various parameters. In all the cases, our proposed framework provided significant improvements over baselines considered for comparison.

## 1. INTRODUCTION

Searching for information is easy when search requirements are well understood and stable. But in those situations where search requirements are not well understood or they are constantly changing, it becomes a challenging task. To help users in such situations, search engines provide various mechanisms to facilitate exploration of search space. Query suggestion is one such mechanism which helps user in query formulation process. It minimizes efforts on the part of the user and leads to early discovery of relevant information. Query suggestion involves suggesting both relevant and related queries as well as diversifying those suggestions. Diversifying suggestions helps in minimizing redundancy and covering various possible search intents of the user.

### 1.1 Motivation

One of the most common reasons for not having well understood and stable information requirements is when there is a sudden change in information needs or when there is sudden availability of new information. Sudden availability of new information can be attributed to new events. When there is a new event, there is a burst in queries related to that topic and information that gets generated in response to that particular event. There is sufficient evidence to support the correlation between event-related content generation and user search behaviour [Anagha Kulkarni et. al. 2011; Yunliang Jiang et. al. 2010; Wisam Dakka et. al. 2012]. This correlation between new events and user search behaviour is more common for news websites, where the user queries have a time sensitive context inherent in them and are seeking event related information most of the time.

Existing work for generating query suggestions [Milad Shokouhi et. al. 2012; Taiki Miyanishi et. al. 2013; Fei Cai et. al. 2014; Stewart Whiting et. al. 2014; Yang Song et. al. 2010; Xiaofei Zhu et. al. 2011] is based on the query logs and user click data. But for news websites, it takes some time before new events get properly represented in query logs. Additionally, due to its dependence on user fired queries, less popular events usually get overshadowed by some highly popular event which invoke greater public interest. For example, immediately after the Pathankot Terrorist Attack in India, when query pathankot was fired, suggestions obtained from Google News did not contain any mention of Pathankot Terrorist Attack (Table 1 First Column). However, when the same query was fired after two months of the incident, even though Pathankot Terrrorist Attack was present in query suggestions, suggestions were almost identical, there was not much diversity in query suggestions (Table 1 Second Column).

On the other hand, using news metadata, we can get suggestions related to Pathankot Attack even in the immediate aftermath of the attack (Table 2 First Column). We can observe from the event centric query suggestions that they are not only relevant but more diverse as well. For example, in second column of Table 1, query suggestions like, Pathankot Attack News, Pathankot Terrorist Attack, Pathankot Attack Video and Pathankot Terror Attack convey almost the same meaning and information, but query suggestions given in second column of Table 2 like, India Pakistan Relation, SIT Team in Pathankot, India Pakistan Match, Pathankot Attack Probe, Jaish e Mohammad, Joint Investigation Team Probe and Pathankot Terror Attack, all of them convey different developments in relation to the Pathankot Terrorist Attack and hence convey different information and search intents. In addition to utilizing the textual similarity of the user query, we can also generate suggestions using event relatedness of two or more events, as event clusters obtained from news metadata behave like nodes in a knowledge graph of events. In Table 2 this unique aspect of event centric query suggestions can be observed from suggestions like, Alert Issued by Delhi Police (First Column), India Pakistan Relations (Second Column) and India Pakistan Match (Second Column), because they are events different from Pathankot Terrorist Attack, but their development was affected by Pathankot Attack.

Table 1. Google Autocomplete Query Suggestions for query - pathankot

| Query Suggestions | |
|---|---|
| **Immediately after Event** | **Two months after Event** |
| Pathankot News | Pathankot Attack |
| Pathankot Airport | Pathankot News |
| Pathankot Weather | Pathankoty Attack News |
| Pathankot Cantt | Pathankot Terrorist Attack |
| Pathankot Punjab | Pathankot SP |
| Pathankot Express | Pathankot Attack Video |
| Pathankot to Delhi Train | Pathankot Terror Attack |
| Pathankot Tourism | Pathankot Air Base |

Table 1. Query Suggestion given by proposed framework for query – pathankot

| Query Suggestions | |
|---|---|
| **Immediately after Event** | **Two months after Event** |
| Pathankot Air Force Base | India Pakistan Relations |
| Pathankot Railway Station | SIT Team on Pathankot |
| Pathankot Terror Attack | Pathankot Attack |
| Pathankot Attack | India Pakistan Match |
| Pathankot Air Force Base Attacks | Pathankot Attack Probe |
| Alert Issued By Delhi Police | Jaish e Mohammad |
| Pathankot Airbase Attack | Joint Investigation Team Probe |
| Terrorist Attack in Pathankot | Pathankot Terror Attack |

Table 3 shows a comparison between query suggestions generated by Google News and actual keywords extracted from news media for various events. Here, we can see that query suggestions generated by Google News are irrelevant in the context of news event taking place. But, news keywords extracted from news media are quite relevant and indicative of the event actually taking place.

Table 3. Comparison of Query Suggestions and News Keywords for News Events

| News Event | Query Suggestion (Google News) | Relevant News Keywords |
|---|---|---|
| Canada Election Results | Canada insurance<br>Canada immigration<br>Canada Express Entry | Canada Polls<br>Liberal Party<br>Indo-Canadian MPs |
| Chhota Rajan Arrest | Chhota Bheem<br>Chhota Rajan<br>Chhota Bheem Video | Chhota Rajan<br>Indonesia<br>Underworld Don |
| Punjab Riots 2015 | Punjab News<br>Punjab Kesari<br>Punjab National Bank | Guru Granth Sahib<br>Guru Granth Sahib Burning<br>Punjab Holy Book Sacrilege |

To summarize, existing query log based suggestion mechanisms are unable to generate event specific suggestions for new events and poorly represent less popular events. Our proposed framework for generating event centric query suggestions solves these two problems using news metadata. For this purpose, we utilize the burst in news media generated content when some new event takes place. This burst of event related information is observed in both user queries related to that event as well as related information generated by news media [Anagha Kulkarni et. al. 2011; Yunliang Jiang et. al. 2010]. As user queries are biased towards popular events, trying to detect such burst of information from news media can provide a more effective solution as news media content is representative of all the events taking place. Metadata of news articles has become more structured and informative over the past few years. Title, description, news keywords, named entities and other such attributes of news articles can be extracted from metadata itself. Therefore, using news metadata suffices to consider various attributes of news events, required by proposed framework to generate query suggestions.

### 1.2 Problem Statement

<u>Event centric Query Suggestion:</u> Event centric query suggestion refers to an ordered list of $n$ news keywords $\langle \alpha 1, \alpha 2, \alpha 3, \dots, \alpha n \rangle$ for a user search query such that at most initial $k$ $(k \leq n)$ news keywords in the list are taken from day wise event clusters and the remaining $n - k$ $(n - k \geq 0)$ news keywords in the list are taken from duration wise event clusters.

To keep the entire process scalable and efficient, we restrict ourselves to metadata of the news articles (Title, Description, News Keywords and Publishing Date). This way we can avoid the expensive task of analyzing the entire news articles for event clustering, keyword extraction and named entities recognition.

### 1.3 Approach Overview

To recommend event centric query suggestions for input query, we first of all create a repository of news keywords and their metadata. Then we group news articles into Day wise and Duration wise event clusters. Day wise event clusters represent news events taking place on a particular day and Duration wise event clusters represent events in their entirety which may span over a number of days. News keywords are ranked inside event clusters and these ranks are used to order candidate suggestions in final query suggestion list. In order to accommodate for both recent events and major events, we mix suggestions from Day wise event clusters and Duration wise event clusters according to mix factor.

### 1.4 Technical Contribution

We make following major technical contributions in the direction of generating event centric query suggestions which are independent of query logs:
   a. Developing a query suggestion framework based on metadata of news articles, which has not been attempted before.
   b. Efficient way of extracting news keywords from metadata and building a repository of news keywords (over 5.5 million news keywords).
   c. A two level clustering algorithm for grouping news events day wise and duration wise, using news metadata.
   d. Ranking mechanism for ranking news keywords within event clusters, which can be used for ordering event clusters and candidate query suggestions.

## 2. RELATED WORK

In this section, we present existing related work which justifies the use of news media content for generating event centric query suggestions. Afterwards, we provide a brief discussion of existing work in the direction of query suggestions. Finally we conclude this section by highlighting limitations of existing works.

### 2.1 Primacy of News Media for Event Related Queries

In [Yunliang Jiang et. al. 2010], Yunliang and others have found that, news and blog articles cover most of the topics resulting out of sudden burst of information about new events. So, using blog or news articles should be sufficient for the purpose of making query suggestions with respect to new events. Another important finding is that query can be better predicted by news content, than by blog content. Authors have suggested that it may be because, many of the search queries are initiated after reading news articles. So building a repository of keywords and metadata extracted from news media is a better choice than using blog metadata and keywords for query suggestions. Within news and blog context, title of the article is more relevant to the query context of time sensitive queries (which are usually result of some new event), than body of the article as title provides a compact representation of the article. It justifies using only metadata extracted information and not analyzing the article body for making event centric query suggestions.

### 2.2 Query Suggestion

Even though there is a technical difference between query auto completion and query suggestion, we have considered existing related work for both of these techniques because both of them focus on generating candidate queries to facilitate user search activities.

**Time Sensitive Query Suggestion:** Time sensitive query suggestion techniques incorporate temporal nature of user queries in addition to their overall popularity while making suggestions. The temporal nature of user queries may arise due to some new events, when there is sudden rise in queries related to a new event. It may also arise due to temporal pattern in user search behavior, like in the morning, users usually search for news or finance related information, while at night, they prefer entertainment related information.

In [Anagha Kulkarni et. al. 2011], authors have considered the problem of understanding temporal queries in term of general features characterizing them and how they evolve over time. In [Wisam Dakka et. al. 2012], authors have estimated relevant time interval for answering time sensitive queries using various techniques like, direct estimation using distribution of matching documents, estimation using binning and estimation using word tracking. Then they have shown how to integrate this temporal relevance in language models, probabilistic models and pseudo relevance techniques for search activities involving time sensitive queries. Raneiri Baraglia and others [Ranieri Baraglia et. al. 2009] have studied the effect of time on query suggestions generated using query flow graphs. They have considered the aging of query flow graphs and hence change in query popularity and user interests with time. TaSQS [Taiki Miyanishi and Tetsuya Sakai 2013], proposed by Taiki and Tetsuya clusters query suggestions along timeline and presents query suggestions after organizing them along timeline. Yang and Li-wei [Yang et. al. 2010] have proposed a query suggestion technique based on implicit user feedback for rare and less popular queries. Here they have considered clicked and skipped urls differently for they contain different level of information with respect to user query. They have proposed a random walk based model. In [Xiaofei Zhu et. al. 2011], Xiaofei Zhu and others have considered query space to be manifold in contrast with the normal practice of considering it as euclidean. They have recommended manifold ranking with stop points for generating both diversified and relevant query suggestions.

**Query Auto Complete Suggestion:** In case of query auto completion, query auto complete suggestions are presented to the user while the query is being typed. In addition to generating meaningful suggestions, query auto completion systems try to minimize the number of user key strokes as well.

Saul and others [Saul Vargas et. al. 2016] propose a term by term query auto completion technique where instead of suggesting entire query to the user, next possible terms for user query are suggested in a step by step manner for mobile devices. In terms of answering time sensitive queries, [Milad Shokouhi and Kira Radinsky 2012], Milad and Kira have used a time series based model and ranked candidate queries on the basis of forecasted frequencies. Stewart and Joemon [Stewart Whiting et. al. 2014] have developed query auto completion ranking approaches based on sliding window, query popularity distribution and short range query popularity prediction. They have tried to accommodate those not yet popular queries which refer to ongoing and emerging events in the list of candidate queries suggested as query auto completion. Fei Cai and Maarten de Rijke [Fei Cai and Maarten de Rijke 2014] have attempted to personalize query auto complete suggestions for time sensitive queries.

**Query Log Independent Query Suggestion:** All of the above works for generating query suggestions are based on analysing user click data available in the form of query logs. In [JH Wang and MH Shih 2015; Sumit Bhatia et. al. 2011], authors have proposed techniques for generating query suggestions without using query logs. In [JH Wang and MH Shih 2015], JH Wang and MH Shih have suggested pseudo relevance feedback based mechanism for term suggestion using query context knowledge bases. They have constructed such knowledge bases using search results extracted from blogs, news and keyterms. In [Sumit Bhatia et. al. 2011], Sumit and others have suggested a probabilistic model that makes use of corpus without depending on query logs. They utilize document corpus for extracting candidate phrases which can be recommended as suggestions for user input query.

### 2.3 Limitation

   a. Despite considerable online presence of news related content contributed by news media, blogs, social media etc., there is lack of query suggestion work which specifically targets information needs of users visiting online news sources.
   b. Majority of existing query suggestion work is based on query log analysis, but query logs have their own limitations in the context of online news media (as discussed in section 1.1). Therefore an effective query suggestion framework for online news should be independent of query logs.

### 3. PROPOSED QUERY SUGGESTION FRAMEWORK

Given an input query $q$, our system generates top $n$ query recommendations. In our work, recommended queries are news keywords. Thus, our first module (metadata extraction and preprocessing) focuses on extracting news keywords from news articles and preprocessing them to build a repository of news keywords. Each news article is considered as an instance of a news event and our purpose is to find out some news events on the daily basis as well as on duration basis. Therefore, news keywords collected by first module are grouped into Day wise and Duration wise event clusters using Clustering and Ranking module. These keywords are ranked inside event clusters and ranks assigned to them are used to order candidate suggestions in final query suggestion list. In order to accommodate for both recent events and major events, query suggestion module mixes suggestions from day wise event clusters and duration wise event clusters according to mix factor, $k$ $(0 \leq k \leq n)$. Figure 1 shows various modules comprising our system.

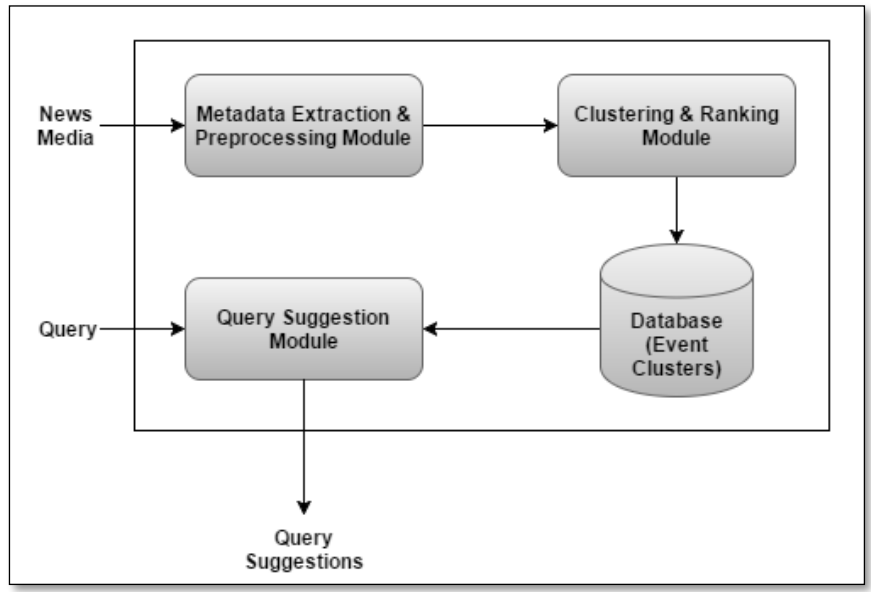

Fig. 1. Block Diagram of Proposed Query Suggestion Framework

### 3.1 Metadata Extraction & Preprocessing Module

For the purpose of creating a repository of news keywords, we crawl data (i.e., news articles) from about 120 different news sources. As for extracting news keywords efficiently from news articles, we experimented with three different techniques – (i) HTML tag based Keyword Extraction, (ii) Heuristic based Keyword Extraction and (iii) Metadata based Keyword Extraction respectively. Of these three techniques, Metadata based Keyword Extraction is most efficient. During metadata preprocessing, we remove noisy and nonessential information which is not required for query suggestion. We perform keywords cleaning, deduplication, and some other processing tasks on extracted keywords. At the end of Preprocessing, each news article is converted into a common and consistent representation of five attributes on which Clustering and Ranking module can operate. These five attribute are – title, description, publishing date, news keywords, and named entities. Metadata Extraction and Preprocessing is described in detail in Section 4.

### 3.2 Clustering and Keyword Ranking Module

Title, Description, Date and News Keywords are used to define feature vector for event clustering. Each news keyword is grouped into Day wise and Duration wise event clusters. Day wise event clusters represent various news events taking place on a particular day and Duration wise event clusters represent major news event in their entirety spanning over multiple days. Day wise event clusters account for recency in query suggestions and Duration wise event clusters account for representation of major news events in query suggestions. Once event clusters are formed, news keywords are ranked inside each event cluster. Ordering of news keywords in event clusters and final query suggestion list is decided by their ranks in event clusters. Named Entities are used for ranking of news keywords within event clusters. Clustering and News Keyword Ranking are explained in detail in Section 5.1 and Section 5.2. Once we have clustered events day wise and duration wise, we index cluster details using lucene [Apache Software Foundation 2015] indexing. Indexing is explained in detail in Section 5.3.

### 3.3 Query Suggestion

When user query is given as input to the system, we utilize full text search functionality of lucene [Apache Software Foundation 2015] to get candidate Day wise and Duration wise event clusters. From these clusters, we extract candidate suggestions for user query. We use the rank assigned to each keyword in Section 5.1 and Section 5.2, and some other parameters like, start date of event, end date of event and weight of the event cluster itself and mix factor, to get the final query suggestion list. Proportion of news keywords from Day wise and Duration wise event clusters is decided in accordance with mix factor, $k$ ($0 \leq k \leq n$). Query suggestion algorithm is explained in detail in Section 5.3.

## 4. METADATA EXTRACTION AND PREPROCESSING

Recent developments in metadata have made it more structured and descriptive of the article content. Open Graph Protocol (OGP) [Facebook 2011] and "news_keywords" tag are two such developments.

In this Section, we first of all, properly introduce metadata of news webpages in the context of these two developments and various attributes of metadata used by our framework. Then, we explain Metadata based Keyword Extraction and Metadata Preprocessing.

### 4.1 Metadata of News Webpages

Metadata refers to data about data. It provides information about the webpage. Metadata content is not displayed as part of the webpage, but its content can be parsed by machines, crawlers, web browsers and other related web services, to get information about the content of the webpage. Metadata content of a typical news webpage contains information like, *description, keywords, author name, last modified date, etc.* It is always in the form of key-value pairs. Figure 2 shows the metadata of a typical news webpage.

```
<meta name="news_keywords" content="Vijay Mallya ,corporate crime, economic offence/ tax evasion">
<meta rel="link" id="http://www.thehindu.com/news/national/sc-directs-vijay-mallya-to-furnish-details/article8524071.ece">
<meta property="fb:app_id" content="137450626398044">
<meta property="og:type" content="article">
<meta property="og:url" content="http://www.thehindu.com/news/national/sc-directs-vijay-mallya-to-furnish-details/article85240
<meta property="og:site_name" content="The Hindu">
<meta property="og:title" content="SC directs Mallya to furnish overseas assets details ">
<meta property="og:description" content="The Supreme Court on Tuesday refused Vijay Mallya's plea to keep his overseas assets
ordered his assets to be revealed to banks for recovery of debts.
"We will even">
<meta property="og:image:type" content="image/jpeg">
<meta property="og:image" content="http://www.thehindu.com/multimedia/dynamic/02385/MALLYA_2385791c.jpg">
<meta name="twitter:image:src" content="http://www.thehindu.com/multimedia/dynamic/02385/MALLYA_2385791c.jpg">
<meta name="DC.date.issued" content="2016-04-26">
<meta property="article:section" content="National">
<meta property="article:published_time" content="2016-04-26T16:23:14+05:30">
```

**Fig. 2 Metadata of a typical News Webpage**

**Open Graph Protocol**
Open Graph Protocol (OGP) [Facebook 2011] provided by Facebook, allows the embedding of web content as Facebook social graph objects. It provides tags which can be used by web content generators to facilitate conversion of web objects into corresponding graph objects. Some of the most frequently used OGP tags are:
    a.  og:type – It describes the type of the content published, like article, audio, video, etc.
    b.  og:url – Its content is link to the webpage.
    c.  og:site_name – It contains name of the web source as provided by the author of the content.
    d.  og:title – Title of the article.

   e.   og:description -  Small description of the webpage content.
   f.   og:image – Url of the image which will be displayed in the resultant social graph object.

News Media has widely adopted these tags to increase their online presence. These OGP tags (along with tags like news_keywords, keywords and article:published_time) can be leveraged to devise an algorithm to extract event related concise but complete information without parsing the entire article text.

**"news_keywords" Tag**
News_Keywords tag allows web content generators to specify which keywords are most relevant to their articles. This tag is widely used by news media to increase the chances of the content getting listed as result for search engine queries. News keywords are quite informative and after some pre-processing, they can be successfully used for the purpose of event clustering and event evolution analysis.

Of various attributes contained in metadata of news articles, we make use of following four attributes (Figure 2), for defining feature vectors for news articles in event space and making event centric query suggestions:
   a.   news_keywords/keywords: Keywords which are central to the theme of the article.
   b.   og:title: Title of the news article.
   c.    og:description: Small summary of the news article.
   d.   article:published_time: Date on which article is published.

### 4.2 Metadata based Keyword Extraction

All the tags/keywords which are assigned to the news article are present as part of news_keywords and/or keywords tag (Figure 3). Other attributes of news articles like title, description and date used by our system can also be obtained from metadata of the news webpages. So, we restricted crawling to head section of the HTML of news articles only (metadata is always present in head section), leaving the main body of the HTML untouched. It significantly reduced the amount of news data crawled and crawling time.

```
<meta name="news_keywords" content="Sania Mirza, Martina Hingis, WTA Finals women's doub
<meta rel="link" id="http://www.thehindu.com/sport/tennis/sania-mirzamartina-hingis-clin
<meta property="fb:app_id" content="137450626398044">
<meta property="og:type" content="article">
<meta property="og:url" content="http://www.thehindu.com/sport/tennis/sania-mirzamartina
article7829713.ece">
<meta property="og:site_name" content="The Hindu">
<meta property="og:title" content="Sania Mirza-Martina Hingis clinch WTA Finals doubles
<meta property="og:description" content="Indian tennis ace Sania Mirza and her Swiss par
```

Fig. 3 "news_keywords" tag and news specific keywords

First of all, head section of the HRML is crawled using Python based crawler. Then, it is parsed using BeautifulSoup package to create DOM object. Meta tags are extracted from the DOM object. From all meta tags, we select the value of those meta tags where the key is either keywords or news_keywords. If both keywords and news_keywords keys are present in meta tags, then union of news keywords specified with both of them is taken. Other attributes like, title, description and date are extracted from values of those meta tags where keys are og:title, og:description and article:published_time respectively. At the end, we have extracted a tuple of the form, ⟨*title, description, keywords, date*⟩ for each news article.

### 4.3 Metadata Preprocessing

Main objective of metadata preprocessing is to process the raw metadata extracted from news media and get rid of noise and all such information which will not be required by the proposed framework. So this module processes the metadata content extracted by crawler and converts each news article into a tuple of five attributes, ⟨*title, description, keywords, date, named entities*⟩.

Metadata preprocessing broadly consists of following three steps:
1. Metadata Deduplication
2. News Keywords Cleaning
3. Named Entities Extraction

#### 4.3.1   Metadata Deduplication

In order to remove duplicate articles crawled from multiple sources on same day, we use 2048 bits sim-hashing [Charikar 2002]. Similarity hashing is used to create fingerprints of the textual documents, which in turn can be used for the purpose of finding out duplicate documents. Algorithms like (md5, sha1) are not very effective at generating fingerprints for textual content because change in a single bit can result in a huge change in hash value. On the other hand, sim-hash is like a compressed version of the text and the hash value generated will be varying a lot, only if the textual content varies a lot. Therefore, it can be used for generating hashes effectively and efficiently.

In order to generate hash value for metadata, title of the article is used as key to the hash function. We then use the hash values to group metadata documents crawled on any particular day into different buckets. Finally to get rid of duplicates, only one metadata document is taken from each bucket.

#### 4.3.2   News Keyword Cleaning

After deduplication, the next task is to remove noise or irrelevant keywords from news keywords present in the metadata of news articles. For removing irrelevant keywords, we apply below mentioned five cleaning steps to keywords extracted from news metadata:

1.  News keywords should contain at least one English character and should not be completely made up of special or numeric characters. This step ensures removal of keywords like, 2016-03-23, 2015, etc. which signify date, year or time.

2.  ***Generic Keyword Removal:*** Many of the keywords specified in metadata are generic keywords or news source name, like *news, gossip, live, indianexpress.com,* etc. These keywords are very generic terms to facilitate content discovery on search engines, and do not provide much information about the news event, like what is the event, where is it taking place, which entities are involved in the event, etc. Therefore, these keywords can be removed without resulting in any loss of news related information.

    Generic keywords have very high frequency of occurrence because they are not news content specific and are tagged with a lot of news articles. This high

frequency of their occurrence can be leveraged for identification of such keywords. After analysing news keywords according to their frequency of occurrence over a period of one week, we identified forty five such keywords (Table 4).

Table 4 Generic News Keywords

| news | articles | bbc | bbc.co.uk |
|---|---|---|---|
| hungama | business standard | toi | daily tribune |
| live | business-standard.com | videos | mydigitalfc |
| fc | mydigitalfc.com | ibnlive | indianexpress |
| latest | indianexpress.com | ft.com | ht48hours |
| updates | financial chronicle | photos | anchorage |
| forecasts | financial times | cnn | highlights |
| streaming | current affairs | abcnews | abc news |
| gossip | tribune india | usatoday | financialtimes |
| photogallery | photo gallery | sunday et | indian express |

3. **URL removal:** Many times, news urls or some other links are present in news keywords, such links are non-informative for our purpose of query suggestion, therefore, such urls are removed in preprocessing step.

4. **Category Name Removal:** Sometimes category names, like *international, cricket, sports, tech,* etc., to which news article belongs are also present as news keywords in metadata. This type of keywords are also not required for query suggestion, therefore we remove them without causing any loss of important keywords.

   In order to identify the category names to which a news article belongs, we take advantage of the structure of news webpage url. Url has a structure which is indicative of the possible category names to which a news article may belong This way of inferring category names for news articles does not require maintaining separate list of category names which has to be created and updated manually.

5. **POS Tagging based Removal:** There are few instances where news sources tokenize the news article title and place the resulting tokens into the *news_keywords* tag. In such scenario, a lot of noise results in news keywords. For example:

   **News Article Title:** *Rajnath in the dark about Pakistan probe team's arrival*
   **Tokenized News Article Title:** *[Rajnath, in, the, dark, about, Pakistan, probe, team, 's, arrival: NN]*
   **News Keywords:** <u>Rajnath</u>, in, the, dark, about, <u>Pakistan</u>, <u>probe</u>, <u>team</u>, 's, <u>arrival</u>
   **POS Tagging:** [*Rajnath: NN, in: IN, the: DT, dark: JJ, about: IN, Pakistan: NNP, probe: NN, team: NNS, 's: POS, arrival: NN*]
   **Final Keywords:** *Rajnath, ~~in~~, ~~the~~, ~~dark~~, ~~about~~, Pakistan, probe, team, ~~'s~~, arrival*

   Here only five keywords are important and rest of the keywords are just noise and we need to get rid of them. To remove such unnecessary noise, we check all *single word keywords* to find if they are noun or verb. If they do not signify a noun or a verb, we remove them *(in, the, dark, about and 's in this case).*

> Though, this type of cases are only few, but since they do occur, we take care of them by Part of Speech (POS) tagging using Stanford CoreNLP [CD Manning et. al. 2014]. As we can see above, only keywords *Rajnath, Pakistan, probe, team* and *arrival* are noun (NN, NNP and NNS signify nouns), therefore they are retained, but remaining keywords are dropped as they constitute noise.

After applying above five keyword cleaning steps, we weed out unnecessary keywords from news keywords extracted from metadata. For example,

<u>Raw News Keywords from Metadata:</u> *Latest News, Nitish Kumar, Lalu Prasad, Sonia Gandhi, Narendra Modi, Sharad Pawar, Bihar News, Battle for Bihar, Swabhimaan rally, National News, Swabhiman Rally in Patna, Politics, National Politics.*

<u>Processed News Keywords:</u> *~~Latest News~~, Nitish Kumar, Lalu Prasad, Sonia Gandhi, Narendra Modi, Sharad Pawar, ~~Bihar News~~, Battle for Bihar, Swabhimaan rally, ~~National News~~, Swabhiman Rally in Patna, ~~Politics~~, ~~National Politics~~.*

#### 4.3.3 Named Entities Extraction

Named Entities Extraction refers to the task of identifying and assigning elements of the text into categories like, *person, organization* and *location*. For the purpose of extracting named entities, we use the content of news_keywords, og:title and og:description tags present in metadata of news articles (Figure 4).

```
<meta name="news_keywords" content="Russia plane crash, Boeing 737-800, FlyDubai plane crash, Rostov crash,W disaster and accident, disaster (general), society, death, transport accident, air and space accident">
<meta rel="link" id="http://www.thehindu.com/news/international/russia-plane-crash/article8373700.ece">
<meta property="fb:app_id" content="137450626398044">
<meta property="og:type" content="article">
<meta property="og:url" content="http://www.thehindu.com/news/international/russia-plane-crash/article837370
<meta property="og:site_name" content="The Hindu">
<meta property="og:title" content="Two Indians among 62 killed in Russia plan crash">
<meta property="og:description" content="All 55 passengers and seven crew of a passenger jet were killed on plane crashed in southern Russia, officials said. The Dubai Media Office says those killed in the crash incl
```

Fig. 4. Metadata of a news article with Named Entities underlined

It has been observed that news keywords which contain named entities are more informative and useful for query suggestion as compared to those keywords which do not contain any named entity. For example, news keywords like, *Russia plane crash, Boeing 737-800, FlyDubai plane crash, Rostov crash* and *Russia* are more informative and useful as compared to keywords like *disaster and accident, society, death, transport accident,* and *air and space accident*. Hence this observation is used while ranking news keywords inside event clusters (News Keyword Ranking is explained in detail in next section). Although named entities are used to rank only news keywords, but we extract them from og:title and og:description tag content as well, because named entity recognition depends on context which is sometimes missing in keywords, but title and description contain complete sentences and hence have context present in them. In order to extract named entities, we make use of StanfordNER Tagger provided by Stanford CoreNLP package [CD Manning et.al. 2014].

Content of og:title, og:description and article:published_time is used as such without any significant preprocessing in the form of title, description and date of the event respectively. So, after we have separated useful information from noise and irrelevant information, we have five attribute representation for each news article – title, description, keywords, date and named entities (Table – 5).

Table 5 Example News Article Representation after Preprocessing Phase

| | |
|---|---|
| **Title** | Two Indians among 62 killed in Russian Plane Crash |
| **Description** | All 55 passenger and seven crew of a passenger jet were killed on Saturday when the plane crashed in southern Russia, officials said. |
| **Keywords** | Russia plane crash, Boeing 737-800, FlyDubai plane crash, Rostov Crash, World, Russia, Disaster and Accident, disaster (general), society, death, transport accident, air and space accident. |
| **Date** | 2016-03-19 |
| **Named Entities** | Russia, Boeing 737-800, FlyDubai, Rostov, Dubai Media Office |

## 5. EVENT CLUSTERING AND KEYWORD RANKING

The five attribute representation of each news article is directly used for first level of clustering (day wise clustering) and keyword ranking (Figure 5). First level clusters and ranked keywords are in turn used for second level of clustering (duration wise clustering) and keyword ranking. Finally day wise and duration wise clusters are indexed using lucene and used for generating candidate query suggestions.

### 5.1 First Level Clustering and Keyword Ranking

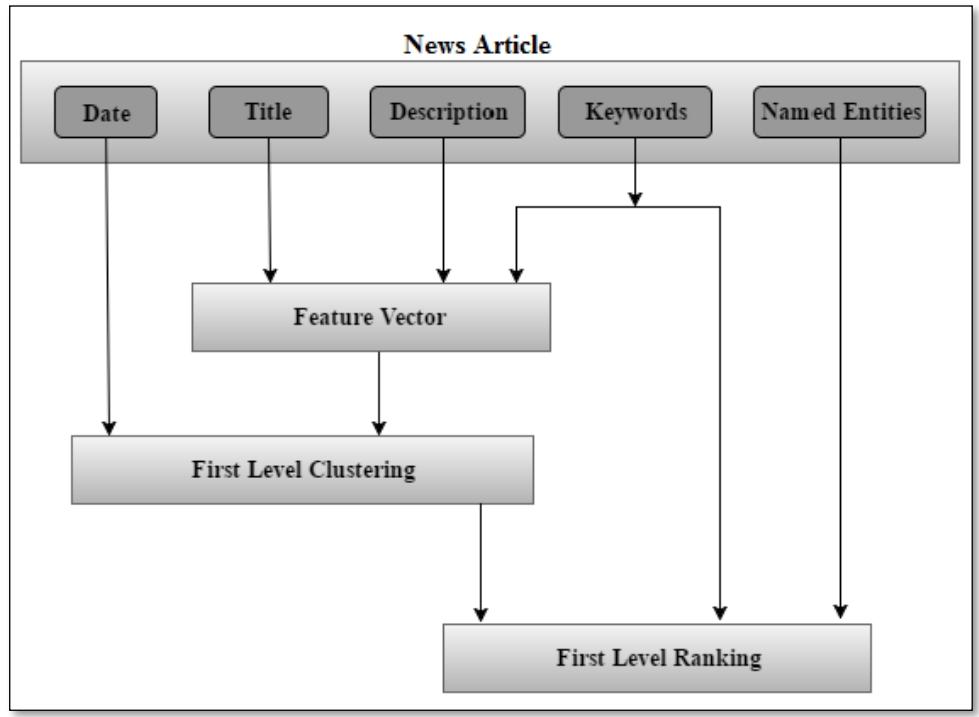

Fig. 5. Block Diagram for First Level Clustering and Ranking

#### 5.1.1 First Level Clustering

First level clustering refers to clustering of the news articles that are published on a particular day into various event clusters. It gives us *day wise event clusters* which can be used to generate query suggestions when a new event takes place and there is sudden burst of event related information. These day wise clusters can be effectively used to generate query suggestions for latest events by clustering various news articles that are published by multiple news sources, because such articles will be similar in content and hence constitute a pattern. On the other hand, query log based approaches even though they manage to generate query suggestions for highly sensational news stories, but they cannot generate suggestions for all latest events, because less popular events usually do not get sufficiently represented in query logs.

First level clustering involves following three sequential sub tasks:
- Feature Vector Creation
- TF-IDF Weighting
- Density based Clustering

**Feature Vector Creation:**
Feature vector refers to n-dimensional vector of features where features are numerical values. It can be used to represent an object. We use *bag of words* representation to define feature vectors for news articles.

Bag of Words Model:
In bag of words representation, each document in a collection of documents is considered as a multiset of its words. Each word is assigned an *id*, and *count* of number of times a particular word appears in the document is recorded. Each document is defined as set of *(id, count)* pairs for each word present inside the dictionary.

Textual content of *title, description* and *keywords* is tokenized using *nltk package* to get unigrams, which are then further reduced to their base forms by *stemming*. Base forms that are obtained after stemming are used as features to get bag of words representation of the news articles.

**TF-IDF Weighting:**
TF-IDF refers to Term Frequency – Inverse Document Frequency. It is a popular weighting scheme for assigning weights to terms (or words) in a document on the basis of how important a term is to a document given a collection of documents. It increases directly with the frequency of occurrence of a term in a documents, but penalizes the occurrence of a term across multiple documents in a collection. Mathematically, it is given by Equation (1).

$$\text{tf-idf}(t, d) = f(t,d) * \log \frac{N}{n(t)} \qquad (1)$$

where, *tf-idf(t, d)* is the tf-idf weight assigned to term *t* in document *d*.
  *f(t,d)* is the frequency of term *t* in document *d*.
  *N* is the total number of documents in the collection.
  *n(t)* is the number of documents in the collection which contain term *t*.

We use TF-IDF weighting to assign weights to various features present in the feature vectors of news articles. Corpus for TF-IDF weighting is defined as entire collection of news articles published on a particular day. It gives us 2D document matrix which is used for density based clustering.

**Density based Clustering:**

We use DBSCAN for clustering news articles published on a particular day to obtain day wise event clusters as DBSCAN can successfully find clusters of arbitrary non-spherical shapes and scales easily to very large sample size.

Density Based Spatial Clustering of Applications with Noise (DBSCAN)
DBSCAN clustering assumes clusters as high density areas which are separated from each other by areas of low density. As an implication of this very generic assumption, it successfully finds clusters of arbitrary shapes. DBSCAN takes as input two parameters: *minSamples* and *eps*. *Core Sample* is the main concept in DBSCAN clustering. A sample is said to be core sample, if there are *minSamples* other samples within a distance of *eps* which are in neighbourhood of it. A sample which is not a core sample is considered as non core sample.

A cluster is a collection of core samples which is created by taking a core sample, finding its neighbours which are core samples and then taking these neighbouring core samples as new core samples and finding their neighbours. This process is repeated until all such core samples are discovered. A cluster may also contain non core samples, if such samples lie in the neighbourhood of a core sample.

For day wise clustering of news articles, we take ***minSamples*** *as **3*** and ***eps*** *as **0.96***.

### 5.1.2 First Level Keyword Ranking

Once news articles have been clustered and assigned as belonging to different events, next task is to rank keywords tagged with those articles on the basis of their importance and usability for the purpose of query suggestion. As stated in Section 4.3.3, news keywords which contain named entities are more useful for query suggestion as compared to those keywords which do not contain named entities. Based on this important observation and others, we came up with a keyword ranking metric for ranking keywords tagged with news articles belonging to a given cluster.

Let, $C_d$ be the set of event clusters obtained on a particular day $d$,
$A_i$ be the set of articles constituting ith cluster $C_{d,i}$,
$A_{i,j}$ be the jth article in article set $A_i$,
$k$ be a keyword tagged with article $A_{i,j}$ in article set $A_i$,
$K$ be the set of tokens (excluding stopwords) constituting keyword $k$,
$title(A_{i,j})$ be the set of tokens (excluding stopwords) constituting title of the article $A_{i,j}$,
$descr(A_{i,j})$ be the set of tokens (excluding stopwords) constituting description of the article $A_{i,j}$.
$NE(A_{i,j})$ be the set of named entities extracted for article $A_{i,j}$.
$n(k)$ be the number of tokens constituting $k$.
$N_{max}$ be the maximum value of $n(k)$ for any keyword k in cluster $C_{d,i}$.

Then, $rank(k, C_{d,i})$ of keyword $k$ in cluster $C_{d,i}$ is given by Equation (2), if $k$ is a named entity otherwise, it is given by Equation (5),

$$\text{rank}(k, C_{d,i}) = \frac{0.1 * n(k)}{N_{max}}, \text{ if } k \in NE(A_i,j), \text{ for some } A_i,j \in A_i \qquad (2)$$

$$\text{rel}(A_i,j) = \text{title}(A_i,j) \cup \text{descr}(A_i,j) \qquad (3)$$

$$\text{rank}(k, A_{i,j}, C_{d,i}) = \left( \sum_{t \in K,\ t \in rel(A_{i,j})} 0.1 \right) + \left( \sum_{t \in K,\ t \in NE(A_{i,j})} 0.1 \right) \quad (4)$$

$$\text{rank}(k, C_{d,i}) = \frac{\left( \sum_{A_{i,j} \in A_i} \text{rank}(k, A_{i,j}, C_{d,i}) \right) * n(k)}{N_{max}} \quad (5)$$

Here, in Equation (2), we keep the rank of a keyword which is also a named entity as constant, irrespective of the number of times it occurs in various articles. The main reason for not incrementing the rank in accordance with its frequency of occurrence is that, even though non named entity keywords may occur in many forms, named entities associated with an event will always be in the same form. As for an example, take *Bihar Election* event, here possible non named entity keywords are *Bihar Polls, Bihar Election, Bihar Elections 2015 and Bihar Assembly Elections,* but named entities like, *Nitish Kumar* will be denoted as *Nitish Kumar* only. Moreover the frequency of occurrence of named entity keywords is usually more than non named entity keywords, therefore, incrementing their rank according to their frequency will lead to giving them very large ranks as compared to non named entity keywords. For non named entity keywords, we increase their ranks on the basis of presence of named entities in them as well as their overlap with content of title and description of article. Keywords with these two characteristics are usually more event specific. In addition to this, we multiply the rank of a keyword by the factor of $\frac{n(k)}{N_{max}}$, because lengthy keywords are more specific in their meaning and are event centric. Table 6 shows some of the day wise clusters obtained on 3rd March, 2016. Table shows keywords along with the ranks assigned to them. Keywords are sorted in decreasing order by their ranks, because keywords with higher ranks will be given priority over keywords with lower ranks while making query suggestions.

Table 6 Example of Day wise Clusters with Keyword Ranks on 03-01-2016

| | |
|---|---|
| Freedom 251 price in india,0.2 | North korea nuclear test,0.6 |
| Freedom 251,0.2 | North korea nuclear,0.375 |
| Ringing bells,0.12 | North korea sanctions,0.375 |
| Freedom 251 price,0.12 | Government and politics,0.15 |
| Freedom 251 specifications,0.12 | Nuclear proliferation sanctions,0.15 |
| Ringing bells refund,0.12 | United nations security council,0.1 |
| Freedom 251 launch,0.12 | United states of america,0.1 |
| Freedom 251 booking,0.12 | Government policy,0.1 |
| Excise duty on gold,0.3 | Super tuesday,0.56 |
| India gold imports,0.225 | 2016 presidential election,0.24 |
| Budget 2016-17,0.2 | Us prez race,0.18 |
| Union budget 2016,0.15 | Us presidential election,0.18 |
| Gold jewellery tax,0.15 | Super tuesday us elections,0.16 |
| Indian gold market,0.15 | Us presidential elections 2016,0.16 |
| Excise duty,0.1 | Republican party leaders,0.12 |
| Gold tax,0.1 | Us presidential elections,0.12 |
| Indian jewellers,0.1 | 2016 united states presidential election,0.1 |
| Jewellers strike,0.1 | |
| India vs Pakistan,0.42 | Controversial hillary clinton emails,0.6 |
| Asia Cup 2016,2.475 | State department releases emails,0.6 |
| Asia Cup T20,0.15 | Hillary clinton super tuesday,0.4 |
| t20 world cup 2016,0.12 | Hillary clinton emails,0.375 |
| 2016 T20 World Cup,0.12 | Hillary clinton state department,0.1 |
| Revenue secretary hasmukh adhia,0.58181 | Apple iphones justice department,1.2 |
| Arun jaitley budget 2016,0.5090 | Apple new york case,0.7 |
| Employee provident fund,0.3545 | San bernardino case,0.6 |
| Provident fund withdrawals,0.2181 | Apple drug case,0.45 |
| Public provident fund,0.1909 | Apple court ruling,0.3 |
| Tax on epf withdrawals,0.1818 | Apple iphone hacking case,0.2 |
| Union budget 2016,0.1636 | Government and politics,0.15 |
| Tax on provident fund,0.1090 | Iphone encryption,0.15 |
| Govt blinks on epf,0.1090 | Computer and data security,0.1 |
| Provident fund,0.1090 | Apple dea,0.1 |
| Arun jaitley,0.1090 | Apple fbi,0.1 |
| Epf tax withdrawal,0.1090 | Locked iphone,0.1 |

### 5.2 Second Level Clustering and Keyword Ranking

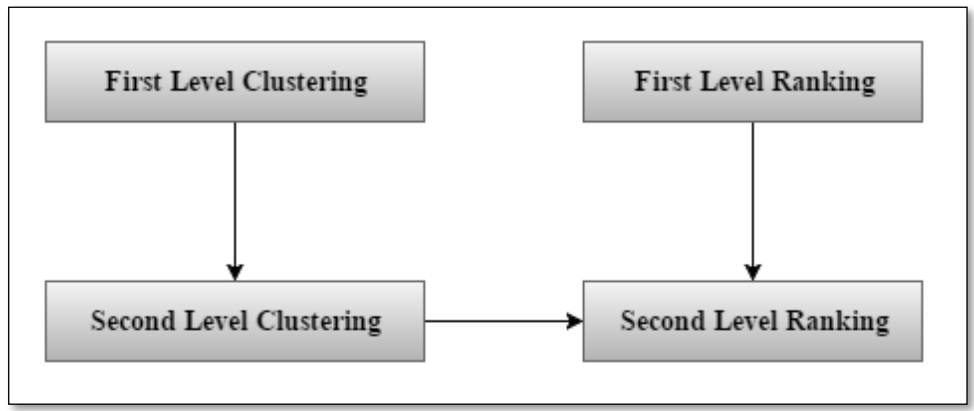

Fig. 6. Block Diagram for Second Level Clustering and Ranking

#### 5.2.1 Second Level Clustering

Second level clustering refers to clustering of day wise clusters that are obtained after clustering news articles published on a particular day (refer Fig. 6). While day wise clusters are aimed at capturing sudden burst of event related data, duration wise clusters are aimed at clustering all day wise clusters for events that span over long duration, like *Bihar Election, Pathankot Terrorist Attack, etc*. This way, they are aimed at summarizing important highlights of such multiday events. It is important to get summarized view of events because not all the events are equally important and do not create same level of impact over a long run, like a *bus accident* or a *robbery* may be relevant and sensational in the immediate aftermath of the event, but in long run, they are not as important an event as *European Migrant Crisis or Paris Attack.* Therefore, summarizing such multiday events allows us to weed out less important and less impacting events over a long period of time using second level of clustering in combination with second level of ranking. In addition to this, certain events are periodic in nature, like *Australian Open, Indian Premier League, Television Reality Shows, etc*. Summarizing events over their entire duration is helpful for such periodic events, as we can suggest important highlights from both recent event as well as its previous version.

Like First level Clustering, it also involves three sequential sub tasks:
- Feature Vector Creation
- TF-IDF Weighting
- Density based Clustering

Second level clustering is done in the same way as First level clustering, only difference is in *feature vector creation* and *parameters for DBSCAN* clustering. In First level clustering, we used the textual content of title, description and keywords as extracted from metadata of news articles for creating feature vectors, but in this case, feature vector is created using *day wise clusters obtained* at the end of First level clustering. For DBSCAN, we set **minSamples** as **2** and **eps** as **0.52**.

### 5.2.2 Second Level Keyword Ranking

Second level Keyword Ranking is done on the top of First level Keyword Ranking. The main purpose of doing Second level ranking is to assign ranks to keywords on the basis of their importance and usability over the entire duration of the event. The cumulative sum of the ranks of individual keywords within a *duration wise cluster* can be used to assign weights to clusters themselves to differentiate more important and sensational events from less important and impacting events. The main reason behind using such aggregated rank score for giving weights to duration wise clusters is that more important and sensational events usually get huge media coverage as compared to less important events and are reported over a long period of time. Therefore the number articles and thereby keywords published for such important events far outnumber the number of articles and keywords published for less important events. For example coverage of events like *Bihar Election, Paris Terror Attack, Chennai Flood* and *Pathankot Terrorist Attack* was far greater than the events like, *Death of Nida Fazli, World Economic Forum Meeting, Anniversary of Peshawar School Massacre* and *Google self-driving Car Accident.*

Let $C$ be the set of duration wise clusters,
$D_i$ be the set of day wise clusters constituting duration wise cluster $C_i$,
$K_{i,j}$ be the set of keywords present in jth day wise cluster $D_{i,j}$ of set $D_i$,
$k$ be some keyword present in $K_{i,j}$.

Then, *rank(k, $C_i$)* of keyword $k$ in duration wise cluster $C_i$ is given by Equation (6).

$$\text{rank}(k, C_i) = \sum_{D_{i,j} \in D_i} \text{rank}(k, D_i, j) \tag{6}$$

Weight of cluster $C_i$, *weight($C_i$)* is given by Equation (7).

$$\text{weight}(C_i) = \sum_{D_{i,j} \in D_i} \sum_{k \in K_{i,j}} \text{rank}(k, C_i) \tag{7}$$

Table 7 Weights of some very popular duration wise clusters

| Event Cluster | Weight of Cluster |
| --- | --- |
| Bihar Election | 368.846932512 |
| Pathankot Terrorist Attack | 237.349404762 |
| Chennai Flood | 144.453690476 |
| Paris Attack | 44.5283333333 |

Table 8 Weights of some less popular duration wise clusters

| Event Cluster | Weight of Cluster |
| --- | --- |
| World Economic Forum Meeting | 7.21 |
| Death of Nida Fazli | 5.075 |
| Google self-driving Car Accident | 2.93333333333 |
| Anniversary of Peshawar School Massacre | 1.233333333 |

In Table 7 and Table 8, we can see a significant difference in the weights of duration wise clusters of some very popular and some less popular events, thereby confirming our observation regarding the extent of media coverage of very sensational and popular events in contrast with less popular events.

Table 9 Top 15 keywords for some popular Duration Wise Clusters

| | |
|---|---|
| Bihar polls,26.5798740149 | Chennai floods,13.8423809524 |
| 2015 Bihar Assembly Elections,22.06436341 | Chennai rains,8.78452380952 |
| Bihar Assembly elections,14.3758100233 | Flood fury in chennai,8.64 |
| Bihar elections,12.5262404262 | Tamil Nadu floods,4.22714285714 |
| 2015 Bihar Assembly polls,9.94641025641 | Chennai flights,3.23 |
| Bihar Assembly polls,9.09151515151 | Chennai floods update,3.15 |
| Bihar election results,6.91296703297 | Chennai weather,2.85642857143 |
| Bihar vidhan sabha election result,4.8076923 | Rescue and relief in chennai,2.8 |
| Second phase voting in Bihar,4.8 | Heavy rain in chennai,2.17142857143 |
| Bihar assembly polls results,3.81538461538 | Chennai airport,2.01333333333 |
| PM Modi in Bihar,3.79761904762 | Tamil Nadu rains,1.85 |
| PM Narendra Modi,3.19899350649 | Chennai rains update,1.575 |
| Bihar Election Counting,2.74615384615 | Modi 1000 crore assistance to Tamil Nadu,1.5 |
| Bihar results,2.74472527472 | Chennai relief operations,1.26857142857 |
| Third phase of Bihar polls,2.14285714286 | Chennai airport flooded, 1.05 |
| Pathankot terror attack,28.1082142857 | Paris attacks,5.76666666667 |
| Pathankot attack,23.4511904762 | Paris terror attacks,3.995 |
| Pathankot Air Force base attack,8.533928571 | France attacks,1.88 |
| Pathankot IAF base attack,5.19619047619 | French envoy to india,1.12 |
| Pathankot air base,4.9775 | Gunmen in paris,1.08 |
| Terror attack in Air Force Base,3.975 | Le bataclan hostage,1.02 |
| Prime Minister Narendra Modi,3.2 | Bataclan attack,0.88 |
| Punjab terror attack,2.81285714286 | Shootout in paris,0.78 |
| Terror attack in India today,2.5 | Terrorist attack in paris,0.64 |
| Terror attack on Pathankot Air Force base,1.85 | Police paris attacks,0.6 |
| National Security Advisor Ajit Doval,1.511904 | Paris restaurant attack,0.54 |
| Lieutenant Colonel Niranjan Kumar,1.428571 | Brussels police detain two,0.5 |
| General Raheel Sharif,1.40571428571 | Paris national stadium attacks,0.48 |
| PM Modi in Pathankot,1.33333333333 | Attack on paris concert hall,0.4 |
| Pathankot attack probe,1.22142857143 | Indians killed in Paris attacks,0.3 |

Table 9 shows some of the top keywords that we get after clustering day wise clusters for four very popular event – *Bihar Election, Chennai Flood, Pathankot Terrorist Attack and Paris Attack.*

Once we are done with First and Second level clustering and keyword ranking, we proceed to indexing the clustering results, so that we can utilize the full text search feature of lucene for implementing the proposed query suggestion framework. Lucene indexing and query suggestion framework are explained in detail in next section.

### 5.3 Query Suggestion using Event Clusters

In order to generate query suggestions using event clusters, *day wise* and *duration wise clusters* are indexed using *lucene*. Full text search feature of lucene is used to discover candidate day wise and duration wise clusters for user query. Afterwards, on the basis of ranks assigned to keywords in candidate clusters, query suggestions are made using a mix of keywords from day wise and duration wise clusters.

**5.3.1  Lucene**

*Lucene* is a software library provided by *Apache* which provides full text indexing and searching functionality. The main feature of lucene is that it treats document as a collection of fields. This assumption regarding the structure of documents makes it independent of file formats. Lucene based indexing and search involves following steps:

1. <u>Build Lucene based Documents</u>: First of all, raw text of the document is organized into various fields of the document. This way text document is stored as collection of fields which lucene based search applications can understand.

2. <u>Analyse Documents</u>: In this step, documents are analysed to decide which fields of the documents should be indexed for searching.

3. <u>Index Documents</u>: Once analysis of documents is over, next step is to index documents. Indexing of the documents keeps track of which phrases of text are present in which documents, thereby speeding up the retrieval of relevant documents while searching.

4. <u>Build Query Object and Searching</u>: User given text query is converted into a format which search mechanism can recognize. Query object so obtained is used to check the index to retrieve relevant documents and other details.

**5.3.2  Event Cluster Documents**

In order to use full search functionality of lucene, we convert event clusters into lucene type documents.

**Day wise Event Cluster Document:** Day wise clusters obtained at the end of First Level Clustering are converted into lucene documents having four fields:

a. <u>Cluster Id:</u> Each cluster on a particular day is uniquely identified by cluster id. It is an *integer* valued field.

b. <u>Keywords:</u> Keywords assigned to a particular event cluster constitute this field. It is a *string* valued field. Full text search is implemented on this field.

c. <u>Cluster Weight:</u> Cluster weight denotes the sum of the ranks of keywords present in the event cluster. It is a *float* valued field. It is one of the parameters used by lucene search to sort search result documents.

d. <u>Date:</u> Date field denotes the date to which event cluster belongs. Date is converted into *timestamp* and stored as *long* valued field. It is also one of the parameters used by lucene search to sort search result documents.

**Duration wise Event Cluster Document:** Duration wise clusters obtained at the end of Second Level Clustering are converted into lucene documents having five fields:

a. <u>Cluster Id:</u> Each event cluster is assigned an id which uniquely identifies it in a set of duration wise event clusters. It is an *integer* valued field.

b. <u>Keywords:</u> Textual content of keywords present in an event cluster constitute this field. It is a *string* valued field. Full text search is implemented on this field.

    c. <u>Cluster Weight:</u> Cluster weight denotes the sum of the ranks of keywords present in the event cluster. It is a *float* valued field. It is one of the parameters used by lucene search to sort search result documents.

    d. <u>Start Date:</u> On the basis of day wise clusters constituting a duration wise cluster, it is assigned an approximate start date. Let, $D$ be the set of day wise clusters constituting a duration wise cluster $c$ and $date(D_i)$ be the date of ith day wise cluster $D_i$ in $D$. Then, start date of duration wise cluster $c$, *start-date(c)* is given by Equation (8).

$$\text{start-date}(c) = \min_{D_i \in D}(\text{date}(D_i)) \qquad (8)$$

Start Date is converted into *timestamp* and stored as *long* valued field. It is one of the parameters used by lucene to sort search results.

    e. <u>End Date:</u> On the basis of day wise clusters constituting a duration wise cluster, it is assigned an approximate end date. Let, $D$ be the set of day wise clusters constituting a duration wise cluster $c$ and $date(D_i)$ be the date of ith day wise cluster $D_i$ in $D$. Then, end date of duration wise cluster $c$, *end-date(c)* is given by Equation (9).

$$\text{end-date}(c) = \max_{D_i \in D}(\text{date}(D_i)) \qquad (9)$$

End Date is converted into *timestamp* and stored as *long* valued field.

### 5.3.3 Indexing

After creating lucene type documents for event clusters, we index them into index database with *Keywords* field set to *analysed*. Setting Keyword field of documents to be analysed allows full text search to be performed on keywords of event clusters. While indexing a new document, we first check if it already exists in the index database. It is indexed only when it does not already exist in the database to avoid duplication.

### 5.3.4 Searching

In order to search the index database, we first of all convert user input query into a query object. Query object is then used to retrieve candidate event clusters using index database. Since we generate suggestions using a mix of both day wise and duration wise clusters, we fetch both day wise and duration wise candidate clusters separately.

Main reason behind using *day wise event clusters* is to accommodate query suggestions from latest events, therefore, while retrieving candidate day wise clusters, they are sorted on the basis of their *date* field in descending order (if two clusters have same *date* field value, then they are sorted in descending order by *cluster weight* field). The purpose of using *duration wise event clusters* is to suggest keywords from major event, thereby, describing events in general and avoiding too specific keywords (which may account for some sub event within main event). Hence, duration wise clusters are sorted on the basis of *cluster weight* field in descending order (if two clusters have same *cluster weight* field value, then they are sorted in descending order by *start date* field).

### 5.3.5 Query Suggestion
**Event centric Query Suggestion:** Event centric query suggestion refers to an ordered list of *n* news keywords ⟨α1, α2, α3, ... , αn⟩ for user search query such that at most initial $k$ ($k \leq n$) news keywords in the list are taken from day wise event clusters and the remaining $n - k$ ($n - k \geq 0$) news keywords in the list are taken from duration wise event clusters.

**Query Suggestion Algorithm:** Query Suggestion algorithm takes three inputs – *user query q, number of query suggestion n* and *mix factor k*. Mix Factor is nothing but an integer value less than or equal to *n*, which decides ratio of keywords from day wise event clusters and keywords from duration wise event clusters. It will in turn decide the share of latest events and major events in the query suggestion list. First *k* suggestions in the output list will be from day wise clusters, i.e., latest events and remaining *n-k* suggestions will be from duration wise clusters, i.e., major events.

---

**ALGORITHM 1.**   Query Suggestion Algorithm

---

**Input:**
 1. User query, **q**
 2. Number of Query Suggestions, **n**
 3. Mix Factor, **k (k≤n)**

**Output:** Ordered list of n news keywords, **R= ⟨α1,α2,α3,…,αn ⟩**.

1. ***Get sorted list of top k day wise clusters, $C1 = \langle \gamma 1, \gamma 2, \gamma 3, \ldots, \gamma k \rangle$ for q.***
2. ***Get sorted list of top l=n-k duration wise clusters, $C2 = \langle \tau 1, \tau 2, \tau 3, \ldots, \tau l \rangle$ for q.***

3. *for $\gamma i \in C1$:*
4.     *sort-desc($\gamma i$)*

5. *for $\tau i \in C2$:*
6.     *sort-desc($\tau i$)*

7.  *max-size-C1* $= \max\limits_{\gamma i \in C1}(size(\gamma i))$
8.  *max-size-C2* $= \max\limits_{\tau i \in C2}(size(\tau i))$

9.  *i = 1*
10. *m = 1*
11. *R = ⟨ ⟩*
12. *while $i \leq$ k and $m \leq$ max-size-C1:*
13.     *j=1*
14.     *while $j \leq size(C1)$:*
15.         *if $i \leq k$ and $m \leq size(\gamma j)$:*
16.             *if $Y_{jm} \notin R$:*
17.                 *add $Y_{jm}$ to R*
18.                 *i = i + 1*
19.         *if i > k:*
20.             *break*
21.         *j = j + 1*
22.     *m = m + 1*

23. *m=1*
24. *while $i \leq n$ and $m \leq$ max-size-C2:*
25.     *j=1*
26.     *while $j \leq size(C2)$:*
27.         *if $i \leq n$ and $m \leq size(\tau j)$:*
28.             *if $\tau_{jm} \notin R$:*
29.                 *add $\tau_{jm}$ to R*
30.                 *i = i + 1*
31.         *if i > n:*
32.             *break*

```
33.              j = j + 1
34.         m = m + 1
35. return R
```

Algorithm 1 shows the pseudo code of the proposed query suggestion algorithm. Here first of all, we fetch the relevant day wise and duration wise clusters for the input query using lucene search. Since lucene search sorts the resulting clusters by *date* for *day wise clusters* and by *cluster weight* for the *duration wise clusters* (refer Section 5.3.2). Steps 1 and 2 give us topmost $k$ day wise clusters and topmost $n$-$k$ duration wise clusters which represent latest event clusters and major event clusters respectively. In steps 3, 4 and steps 5, 6, we sort the keywords present in corresponding day wise and duration wise clusters in descending order by the value of the rank assigned to them in Section 5.1 and Section 5.2. This way, preference is given to those keywords in event clusters which are more useful and important for the purpose of query suggestion. In steps 7 and 8, *max-size-C1* and *max-size-C2* store maximum number of keywords that can be found in any event cluster of $C1$ and $C2$ respectively. In *while loop* of steps 12 to 22, we select topmost keywords from each event cluster of $C1$. In the ideal case of lucene search returning exactly $k$ day wise event clusters, exactly one keyword will be selected from each event cluster of $C1$. But, if the number of event clusters returned is less than $k$, then, first we select topmost keyword from each event cluster and if the number of keywords selected is less than $k$, then we select second topmost keyword from each cluster. This process is repeated until either $k$ keywords have been selected ($i > k$) or we have exhausted all the event clusters ($m > max\text{-}size\text{-}C1$). In the while loop of steps 24 to 34, we select topmost keywords from each event cluster of $C2$. In this case also, we follow the same methodology that we used in case of day wise event clusters. In step 35, final list of query suggestions is returned. Table 10 shows query suggestions generated for various queries with query suggestion size, $n = 8$ and mix factor, $k = 2$ and $k = 4$.

Table 10 Query Suggestions for n = 8 and k = 2, 4

| Query | Query Suggestions, k = 2 | Query Suggestions, k = 4 |
|---|---|---|
| donald trump | south carolina republican primary<br>donald trump campaign<br>donald trump on fox debate<br>donald trump 2016<br>iowa caucuses<br>donald trump town hall<br>donald trump president<br>donald trump unafraid & unashamed tour | south carolina republican primary<br>donald trump campaign<br>us government<br>pope francis says donald trump not christian<br>donald trump on fox debate<br>donald trump 2016<br>iowa caucuses<br>donald trump town hall |
| narendra modi | pm narendra modi<br>narendra modi government<br>bihar polls<br>pathankot terror attack<br>paris climate summit<br>modi uk visit<br>netaji files<br>fire at make in india event | pm narendra modi<br>narendra modi government<br>conspiracies being hatched to destabilise govt<br>modi at bhu<br>bihar polls<br>pathankot terror attack<br>paris climate summit<br>modi uk visit |
| hillary Clinton | south carolina republican primary<br>hillary clinton<br>democratic debate<br>iowa caucuses<br>bernie sanders hillary clinton debate<br>2016 united states presidential election<br>us presidential elections: hillary clinton | south carolina republican primary<br>hillary clinton<br>hillary meets britney<br>hillary clinton 2016<br>democratic debate<br>iowa caucuses<br>bernie sanders hillary clinton debate<br>2016 united states presidential election |

| | democratic presidential candidate hillary Clinton | |

## 6. EXPERIMENTAL RESULTS AND DISCUSSION

### 6.1 Dataset

For the purpose of data collection, we crawled metadata of news webpages from about 120 different news sources across the world. Table 12 lists 24 of various news sources crawled by crawler. All the news sources considered for the experiment had news keywords present in their metadata. Since we need only metadata of news articles, instead of crawling the entire news webpages, we crawled only the head section of the html of news webpages. It drastically reduced crawling time as well as the amount of data crawled. The reduction in crawling time varies for each news article depending on the network congestion at that time, size of the article, news source and many other parameters. So, we measured the reduction in average crawling time over a set of 3,815 news articles and found a reduction in crawling time by 29.92%. To measure the reduction in amount of data crawled when crawling only head section (rather than crawling the entire HTML of the webpage), we measured it over a set of 3,702 news articles. We found a reduction of 90.17% in average amount of data crawled. Table 11 shows general dataset characteristics. For the purpose of making a large enough dataset, we crawled data for about seven months starting from October, 2015 to May, 2016. It gave us a dataset containing 5,571,057 news keywords representing 934,426 news articles. On an average, about 4,209 news articles are crawled per day. After applying two level event clustering as explained in Section 5, a total of 30,614 day wise clusters were obtained, resulting in about 138 day wise clusters per day. Total number of duration wise clusters obtained is 1,805. For first level of event clustering, we have used DBSCAN clustering with minSamples as 3 and eps as 0.96. For second level of event clustering, we have used DBSCAN clustering with minSamples as 2 and eps as 0.52.

Table 11 General Dataset Characteristics

| | |
|---|---|
| Duration of Data Collection | 10/03/2015– 05/14/2016 |
| Total Number of News Keywords in Dataset | 5571057 |
| Total Number of News Articles | 934426 |
| Average Number of News Articles (per day) | 4209.13 |
| Total Number of Day wise Clusters | 30614 |
| Average Number of Day wise Clusters (per day) | 138.5 |
| Total Number of Duration wise Clusters | 1805 |

Table 12 Various News Sources Used for Data Collection

| | | |
|---|---|---|
| India Today | Indian Express | Indian Express |
| The Hindu | Hindustan Times | CNN |
| USA Today | BBC | NDTV News |
| Zee News | IBN Live | ABC News |
| Economic Times | NBC News | Today News |
| Washington Post | Chicago Tribune | Sun Times |
| CNBC | Tribune India | Business Standard |
| TechCrunch | Financial Express | Reuters India |

Figure 7 shows frequency distribution of news articles crawled per day over the period of data collection. Similarly, Figure 8 shows frequency distribution of Day wise Event Clusters formed per day over the same period. In Figure 7 and Figure 8, an alternating pattern of peaks and troughs is visible. A detailed analysis of this variation showed that peaks usually correspond to weekdays and troughs usually correspond to weekdays. On an average, the number of news articles published on weekdays is higher than the number of news articles published on weekends. It consequently leads to higher number of Day wise Event Clusters on weekdays than on weekends.

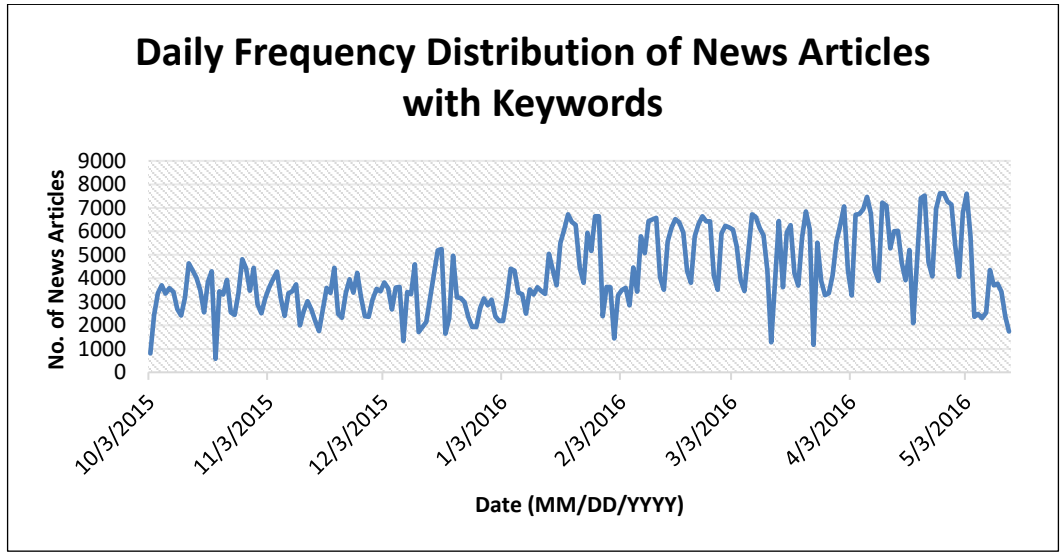

Fig. 7. Daily Frequency Distribution of News Articles (containing keywords)

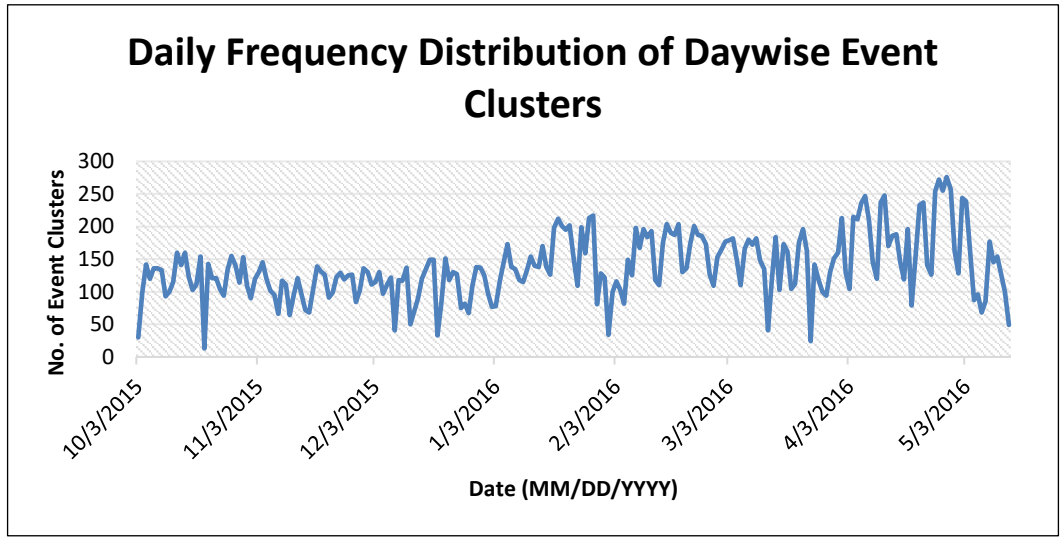

Fig. 8. Daily Frequency Distribution of Day wise Event Clusters

### 6.2 Evaluation of Query Suggestion Framework

#### 6.2.1 Baselines

In order to assess the quality of query suggestions against existing query suggestion mechanisms, we tested proposed framework against four baselines – *Google News, Bing News, Google Search and Bing Search*. *Google News* and *Bing News* belong to the category of very few search engines completely dedicated to online news. *Google Search* and *Bing Search* are some of the most popular and most advanced search engines available for general purpose search activity.

#### 6.2.2    Query Dataset

In order to keep the evaluation objective and free from bias, we created a dataset of *Google News Top Stories*. Google News Top Stories are keywords signifying popular news stories at that particular time. Examples of Google News Top Stories are – Zika Virus, South China Sea, Donald Trump, Samsung Galaxy, etc. These keywords refer to popular events like, Zika Virus Outbreak, Chinese Aggression in South China Sea, US Presidential Election and Launch of Samsung Galaxy respectively. In subsequent discussions, User queries refer to Google News Top Stories for all experimental purposes. Query suggestions generated by baselines and proposed framework for such top stories are used for the comparison of effectiveness of various systems in fulfilling information requirements of users in the context of online news.

#### 6.2.3    Query Suggestion Example

Table 13 and Table 14 shows a comparison of query suggestions for search query – "Narendra Modi", between various baselines and proposed framework at mix factor, k = 2 and 4. It can be seen from the table that query suggestions generated by baselines are quite generic and are not event oriented. Even query suggestions offered by Google News and Bing News which are dedicated web search engines for online news, are missing the event context in their suggestions, whereas getting news event related information is the primary information requirement for which a user visits online news websites. On the other hand, query suggestions generated by proposed framework are event specific and suggest various events reported by news media representing the possible search intents of the user query.

Table 13 Query Suggestion for query – "narendra modi" by Google News, Bing News and Proposed Framework for k=2

| *Google News* | *Bing News* | *Proposed Framework, k=2* |
| --- | --- | --- |
| 1. narendra modi app | 1. narendra modi | 1. pm narendra modi |
| 2. narendra modi twitter | 2. narendra modi app | 2. narendra modi government |
| 3. narendra modi pakistan | 3. narendra modi news | 3. bihar polls |
| 4. narendra modi speech | 4. narendra modi latest news | 4. pathankot terror attack |
| 5. narendra modi latest news | 5. narendra modi biography | 5. paris climate summit |
| 6. narendra modi wikipedia | 6. narendra modi twitter | 6. modi uk visit |
| 7. narendra modi in noida | 7. narendra modi app for windows | 7. netaji files |
| 8. narendra modi wife | 8. narendra modi biography in hindi | 8. fire at make in india event |

Table 14 Query Suggestion for query – "narendra modi" by Google Search, Bing Search and Proposed Framework, k=4

| *Google Search* | *Bing Search* | *Proposed Framework, k=4* |
| --- | --- | --- |
| 1. narendra modi news | 1. narendra modi app | 1. pm narendra modi |
| 2. narendra modi twitter | 2. narendra modi news | 2. narendra modi government |
| 3. narendra modi wife | 3. narendra modi latest news | 3. conspiracies being hatched to destabilise govt |
| 4. narendra modi height | 4. narendra modi biography | 4. modi at bhu |
| 5. narendra modi net worth | 5. narendra modi twitter | 5. bihar polls |
| 6. narendra modi speech | 6. narendra modi wiki | 6. pathankot terror attack |
| 7. narendra modi facebook | 7. narendra modi app for windows | 7. paris climate summit |
| 8. narendra modi latest | 8. narendra modi biography in hindi | 8. modi uk visit |

#### 6.2.4    Diversity Comparison

We measure diversity of query suggestions using a similar procedure as used in [Xioafei Zhu et. al. 2011]. Specifically we measure diversity using difference in the topmost urls presented as search results by a search engine (Google in our case) for queries in a given query suggestion list. Since search engines generally tend to offer different number of query suggestions for different user queries, we sampled out such queries, for which all the four baselines as well as the proposed framework generated at least eight query suggestions. Mathematically, diversity of query suggestion list for a given query q is given by Equation (11).

Let Q be the list of query suggestions generated for user query q, sim($q_1$, $q_2$) be the number of same urls in topmost $n$ search results of query $q_1$ and $q_2$ ($q_1$, $q_2 \in Q$) and $|Q|$ be the size of query suggestion list Q.

Then,

$$d(q_1, q_2) = 1 - \frac{sim(q_1, q_2)}{n} \qquad (10)$$

$$\text{diversity}(q) = \sqrt{\frac{\sum_{q_1 \in Q} \sum_{q_2 \in Q} d(q_1, q_2)}{|Q|(|Q|-1)}} \qquad (11)$$

We measure average diversity using n = 10 and $|Q|$ = 2, 3, 4, 5, 6, 7 and 8 for a set of 50 user queries. Figure 9 shows average diversity values for different query suggestion sizes at mix factor value of 0, i.e., all the suggestions are generated using duration wise clusters. It means we are considering only major events and are not considering latest events explicitly. At mix factor of zero, our proposed framework gives better results than Google News, Bing News and Bing Search, but it performs poorly as compared to Google Search for most of the query suggestion sizes. Hence, it reinforces our assumption that making most effective query suggestions requires making suggestions from both latest events as well as major events. Figure 10 shows average diversity comparison for different query suggestion sizes at mix factor of 2. Here proposed framework clearly outperforms all the four baselines except at N = 2, where its performance is less than that of Google Search. Similarly, Figure 11 shows comparison at mix factor of 4, here also, except for query suggestion size of 2, our system outperforms all the baselines. Figure 12 and Figure 13 also show similar trends at mix factors of 6 and 8 respectively.

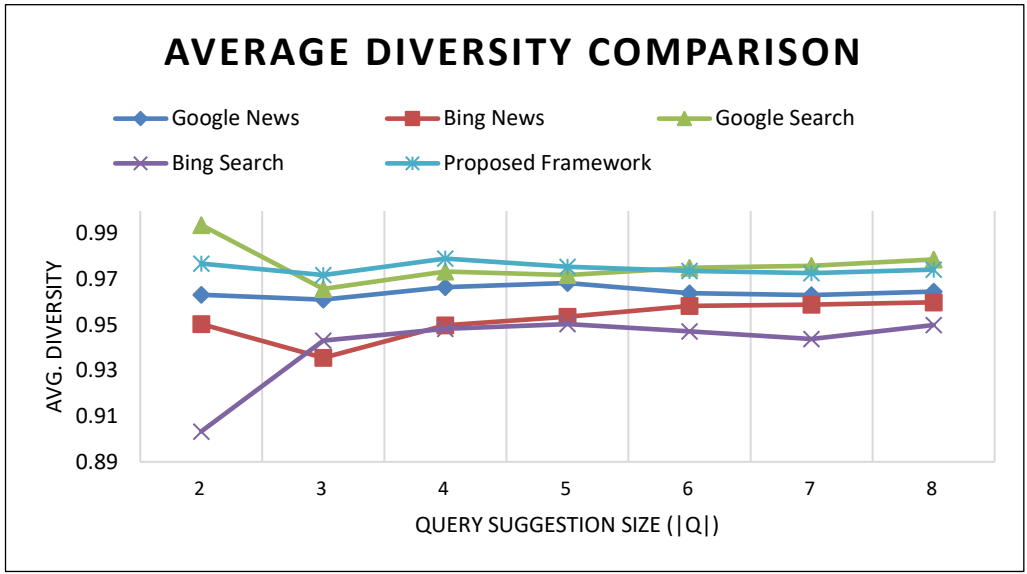

Fig. 9. Average Diversity Comparison for different Query Suggestion Sizes at Mix Factor, k = 0

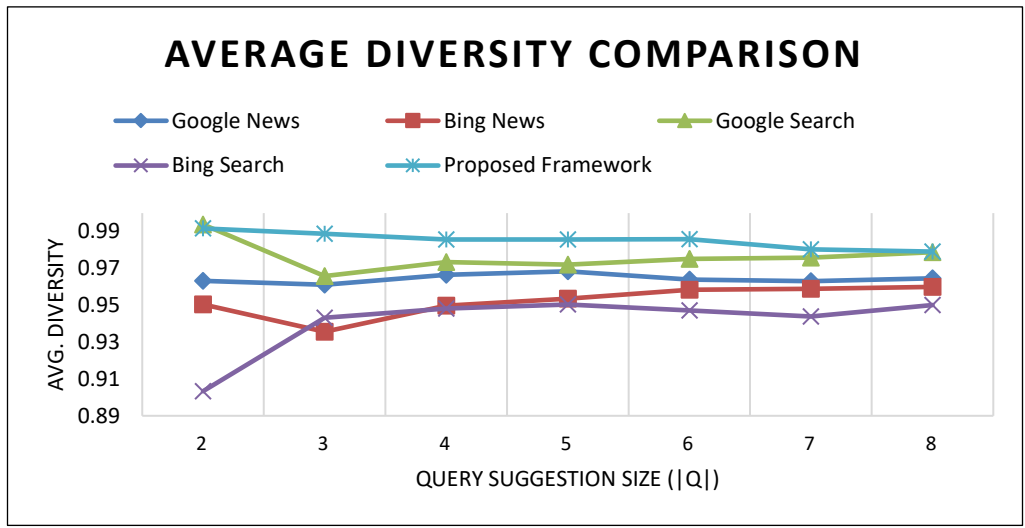

Fig. 10. Average Diversity Comparison for different Query Suggestion Sizes at Mix Factor, k = 2

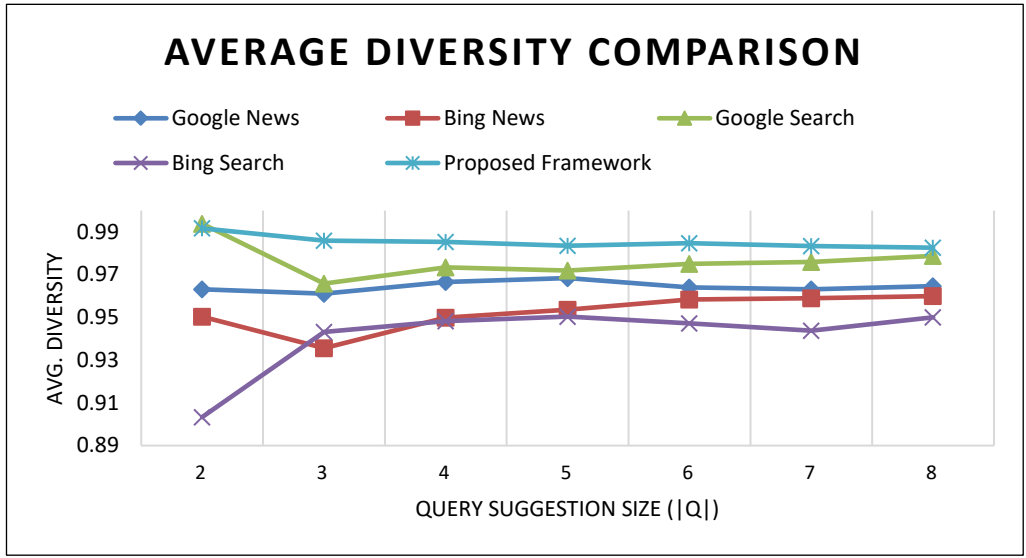

Fig. 11. Average Diversity Comparison for different Query Suggestion Sizes at Mix Factor, k = 4

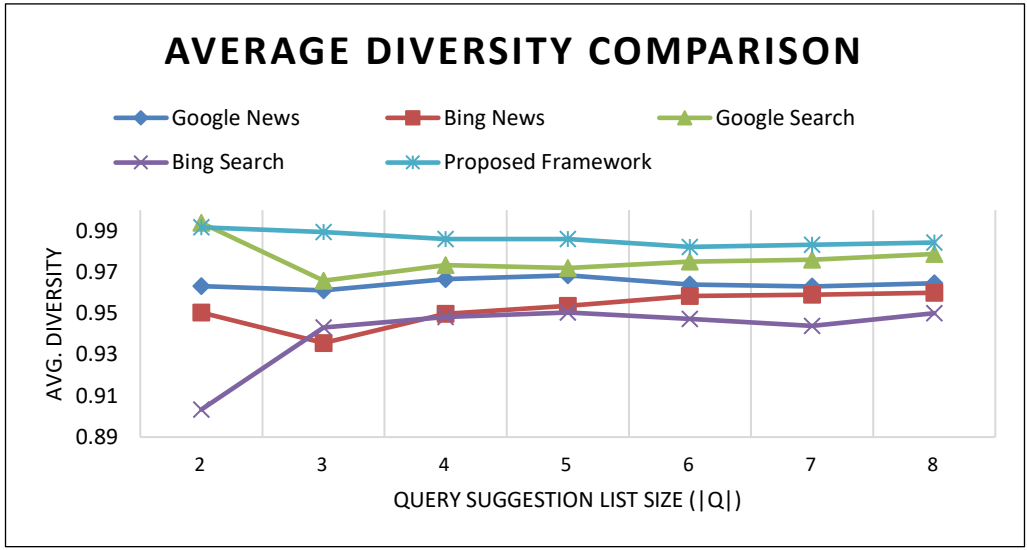

Fig. 12. Average Diversity Comparison for different Query Suggestion Sizes at Mix Factor, k = 6

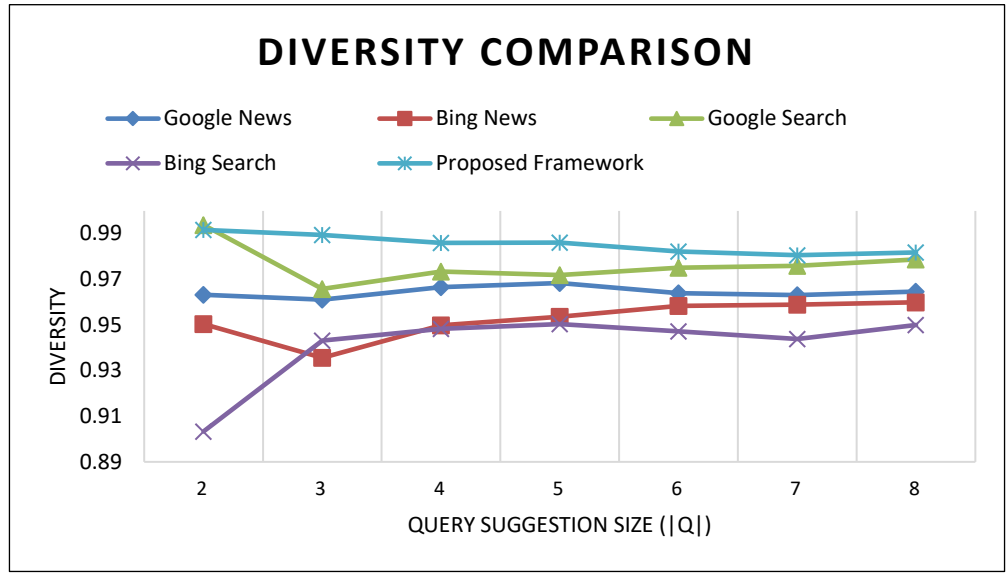

Fig. 13. Diversity Comparison for different Query Suggestion Sizes at Mix Factor, k = 8

### 6.2.5 Effectiveness for New Events

When a new event happens, there is sudden burst of event related information. It takes some time before user search behavior for a particular event gets completely captured in the form of query logs. In addition to this, some events may get overshadowed by other events, leading to their insufficient representation in query logs. Therefore, we also performed a comparison of number of query suggestions that get generated for a particular query.

Table 15 shows a comparison of average size of query suggestion list obtained using different techniques for a set of 50 user queries. We measured number of query suggestions that get generated on two occasions to account for adoption by query log based approaches of new events over time. We initially measured it on 02/19/2016 and again measured it on 02/24/2016. Over a period of five days, number of query suggestions generated by Bing News and Bing Search improved, whereas, no change was detected for all other techniques. Even though, it improved with time for both the two baselines, overall our approach generates more number of query suggestions for new events. Thus our approach can better generate sufficient number of suggestions even when event has recently been discovered without discriminating between highly sensational events and less popular events.

Table 15 Average Query Suggestion Size, initially on 02/19/2016 and finally on 02/24/2016

| Technique | Initial Count Average (Std. Deviation) | Final Count Average (Std. Deviation) |
|---|---|---|
| Google News | 4.04 (3.72469238) | 4.04 (3.72469238) |
| Bing News | 5.60 (3.13581462) | 5.64 (3.28988348) |
| Google Search | 6.68 (2.80950766) | 6.68 (2.80950766) |
| Bing Search | 5.24 (3.35757849) | 5.36 (3.54588963) |
| **Proposed Framework, k=0, 2, 4, 6, 8** | **7.76 (1.01159939)** | **7.76 (1.01159939)** |

#### 6.2.6   User Study

In addition to automatic evaluation of the quality of the query suggestions, we also conducted a user study. For this purpose, we considered a set of 20 user queries and for each user query generated 8 query suggestions. We asked human judges to answer following three questions regarding the quality of query suggestions for each user query from the perspective of online news – *Number of relevant suggestions, Number of unique events that can be inferred from query suggestions* and *Number of surprising but useful suggestions.* For each of the four baselines and proposed framework (with different mix factors), we asked judges to answer each of the three questions on a scale from 0 to 8. In order to keep the user study fair, we intermixed the query suggestion lists generated by various techniques and then asked the users to answer above questions.

Table 16 shows the results of user study for different techniques. The quality of query suggestions generated by our approach is significantly better than that of the baselines on all the three parameters. In contrast to all the previous experiments where query suggestions generated by Google Search provided better results than rest of the baselines, here, Google News gives better results than any other baseline. This may be due to the fact that query suggestions generated by Google News which is dedicated search engine for online news, looked more meaning in the context of online news to human judges. This type of inference cannot be made by any of the preceding automatic evaluation. But the outcomes of our proposed framework clearly outperform even results obtained by Google News.

**Table 16 User Study Results**

| Technique | Relevant Suggestions | Unique Events | Surprising Suggestions |
|---|---|---|---|
| Google News | 2.69375 | 1.45625 | 0.4875 |
| Bing News | 2.05 | 0.95 | 0.2125 |
| Google Search | 1.95 | 0.8125 | 0.1125 |
| Bing Search | 2.075 | 0.75625 | 0.1375 |
| Proposed Framework, k=0 | 5.7 | 3.2875 | 2.7 |
| Proposed Framework, k=2 | 7.15 | 5.76875 | 3.75 |
| Proposed Framework, k=4 | 6.225 | 3.55625 | 2.16875 |
| Proposed Framework, k=6 | 6.5375 | 3.85 | 2.40625 |
| Proposed Framework, k=8 | 6.45625 | 3.74375 | 2.7 |

In all the experiments, our framework clearly performs better than any of the baselines for nonzero value of mix factor. Hence, proving the effectiveness of our approach in comparison with other existing systems. It also reinforces our assumption that in order to make effective query suggestions from the perspective of online news, we need to select news keywords from both latest events as well as overall major events.

### 7.   CONCLUSION AND FUTURE WORK

In this dissertation work, we proposed a novel approach for generating query suggestions targeting information requirements of users visiting online news websites. We suggest an event centric query suggestion framework built on top of news keywords extracted from news metadata. Our approach is independent of query logs and hence it provides a solution for traditional limitations of query logs in context of suggesting relevant suggestions for news events. In order to make suggestions from both latest and major events, we suggest a two level event clustering and two level keyword ranking. Day wise and duration wise clusters obtained at the end of two level clustering can be used successfully for making query suggestions.

In order to evaluate the effectiveness of proposed framework against various baselines, we tested our approach using both automatic evaluation as well as user study. We thoroughly tested the quality of query suggestions against a number of parameters and found our framework gave better results than all the four baselines for nonzero mix factor, thereby suggesting the suitability of event centric query suggestion in facilitating the user search activity on a new website.

In future, we would like to extend this work to find the optimal mix factor value which gives the most effective query suggestions, as our current experimental analysis could not establish the superiority of one particular mix factor value. Also we would like to build a full-fledged system to analyse the user session data and make user context aware query suggestions by add more dimensions to proposed query suggestion framework.

## REFERENCES

Anagha Kulkarni, Jaime Teevan, Krysta M. Svore, Susan T. Dumais. 2011. Understanding temporal query dynamics. *Proceedings of the fourth ACM international conference on Web search and data mining, 2011.*

Yunliang Jiang, Cindy Xide Lin, Qiaozhu Mei, Ann Arbor. 2010. Context comparison of bursty events in web search and online media. *Proceedings of the 2010 Conference on Empirical Methods in Natural Language Processing, 2010.*

Wisam Dakka, Luis Gravano, and Panagiotis G. Ipeirotis. 2012. Answering General Time-Sensitive Queries. *IEEE Transactions on Knowledge and Data Engineering, 2012.*

Facebook. 2011. Open Graph Protocol Specification, *Available at http://ogp.me/, Nov. 2011.*

Apache Software Foundation. 2015. Apache Lucene 5.3.1, *Available at https://lucene.apache.org/.*

Xiaofei Zhu, Jiafeng Guo, Xueqi Cheng, Pan Du, Hua Wei Shen. 2011. A unified framework for recommending diverse and relevant queries. *Proceedings of the 20th international conference on World wide web, 2011.*

Ranieri Baraglia, Carlos Castillo, Debora Donato, Franco Maria, Raffaele Perego, Fabrizio Silvestri. 2009. Aging effects on query flow graphs for query suggestion. *Proceedings of the 18th ACM conference on Information and knowledge management, 2009.*

Milad Shokouhi, Kira Radinsky. 2012. Time-Sensitive Query Auto-Completion. *Proceedings of the 35th international ACM SIGIR conference on Research and development in information retrieval, 2012.*

Taiki Miyanishi, Tetsuya Sakai. 2013. Time-aware Structured Query Suggestion. *Proceedings of the 36th international ACM SIGIR conference on Research and development in information retrieval, 2013.*

Fei Cai, Maarten de Rijke. 2014. Time-sensitive Personalized Query Auto-Completion. *Proceedings of the 23rd ACM International Conference on Conference on Information and Knowledge Management, 2014.*

JH Wang, MH Shih. 2015. Constructing Query Context Knowledge Bases for Relevant Term Suggestion. *Journal of Information Science and Engineering, 2015.*

Stewart Whiting, Joemon M. Jose. 2014. Recent and robust query auto-completion. *Proceedings of the 23rd international conference on World Wide Web, 2014.*

Yang Song, Li-wei He. 2010. Optimal rare query suggestion with implicit user feedback. *Proceedings of the 19th international conference on World Wide Web, 2010.*

Sumit Bhatia, Debapriyo Majumdar, Prasenjit Mitra. 2011. Query suggestions in the absence of query logs. *Proceedings of the 34th international ACM SIGIR conference on Research and development in Information Retrieval, 2011.*

Saúl Vargas, Roi Blanco, Peter Mika. 2016. Term-by-Term Query Auto-Completion for Mobile Search. *Proceedings of the Ninth ACM International Conference on Web Search and Data Mining, 2016.*

MS Charikar. 2002. Similarity Estimation Techniques from Rounding Algorithms. *STOC '02 Proceedings of the thirty fourth annual ACM symposium on Theory of computing, 2002.*

CD Manning, Mihai Surdeanu, John Bauer, Jenny Finkel, Steven J. Bethard, David McClosky. 2014. The Stanford CoreNLP Natural Language Processing Toolkit. *Proceedings of the 52nd Annual Meeting of the Association for Computational Linguistics: System Demonstrations, 2014.*